\documentclass[letter]{aa}  

\usepackage{float}
\usepackage{graphicx}
\usepackage{txfonts}
\usepackage[colorlinks = true,
            linkcolor = blue,
            urlcolor  = blue,
            citecolor = blue,
            anchorcolor = blue]{hyperref}
\usepackage{verbatim}

\newcommand{\msun}{{\ensuremath{M_{\odot}}}}
\newcommand{\mprim}{\ensuremath{M_{\mathrm{prim}}}}
\newcommand{\mcomp}{\ensuremath{M_{\mathrm{comp}}}}

\newcommand{\rsun}{\ensuremath{R_{\odot}}}

\newcommand{\kms}{\ensuremath{\mathrm{km}\,\mathrm{s}^{-1}}}

\newcommand{\dday}{\ensuremath{\mathrm{d}}}

\newcommand{\gauss}{\mathrm{G}}

\newcommand{\alfven}{Alfv\'en\@\xspace}
\newcommand*{\eg}{e.g.\@\xspace}
\newcommand*{\ie}{i.e.\@\xspace}
\newcommand*{\cf}{cf.\@\xspace}
\newcommand{\mesa}{\mbox{\textsc{mesa}}\xspace}
\newcommand{\arepo}{\mbox{\textsc{arepo}}\xspace}

\begin{document} 

\title{Bipolar planetary nebulae from common envelope evolution of binary stars}

\author{Patrick A. Ondratschek\inst{1,3}
    \and Friedrich K. R{\"o}pke \inst{2,3}
    \and Fabian R.N. Schneider \inst{3,4}
    \and Christian Fendt \inst{5}
    \and Christian Sand \inst{3}
    \and Sebastian T. Ohlmann \inst{6}
    \and R{\"u}diger Pakmor \inst{7}
    \and Volker Springel \inst{7}
}

\institute{
    Max Planck Institute for Solar System Research, Justus-von-Liebig-Weg 3, 37077 Göttingen, Germany\\
    \email{ondratschek@mps.mpg.de}
    \and Zentrum f\"{u}r Astronomie der Universit\"{a}t Heidelberg, Institut f{\"u}r Theoretische Astrophysik, Philosophenweg 12, 69120 Heidelberg, Germany
    \and Heidelberger Institut f\"{u}r Theoretische Studien, Schloss-Wolfsbrunnenweg 35, 69118 Heidelberg, Germany
    \and Zentrum f{\"u}r Astronomie der Universit{\"a}t Heidelberg, Astronomisches Rechen-Institut, M{\"o}nchhofstr.\ 12-14, 69120 Heidelberg, Germany
    \and Max Planck Institute for Astronomy, K\"onigstuhl 17, 69117 Heidelberg, Germany
    \and Max Planck Computing and Data Facility, Gießenbachstraße 2, 85748 Garching, Germany
    \and Max Planck Institute for Astrophysics, Karl-Schwarzschild-Str.\ 1, 85748 Garching, Germany
}
             
\date{Received YYY; accepted ZZZ}

\abstract{

Asymmetric shapes and evidence for binary central stars suggest a common-envelope origin for many bipolar planetary nebulae. The bipolar components of the nebulae are observed to expand faster than the rest and the more slowly expanding material has been associated with the bulk of the envelope ejected during the common-envelope phase of a stellar binary system. Common-envelope evolution in general remains one of the biggest uncertainties in binary star evolution and the origin of the fast outflow has not been explained satisfactorily. We perform three-dimensional magnetohydrodynamic simulations of common-envelope interaction with the moving-mesh code \arepo. Starting from the plunge-in of the companion into the envelope of an asymptotic giant branch star and covering hundreds of orbits of the binary star system, we are able to follow the evolution to complete envelope ejection. We find that magnetic fields are strongly amplified in two consecutive episodes. First, when the companion spirals in the envelope and, second, when it forms a contact binary with the core of the former giant star. In the second episode, a magnetically-driven, high-velocity outflow of gas is launched self-consistently in our simulations. The outflow is bipolar and the gas is additionally collimated by the ejected common envelope. The resulting structure reproduces typical morphologies and velocities observed in young planetary nebulae. We propose that the magnetic driving mechanism is a universal consequence of common envelope interaction responsible for a substantial fraction of observed planetary nebulae. Such a mechanism likely also exists in the common-envelope phase of other binary stars that lead to the formation of Type Ia supernovae, X-ray binaries and gravitational-wave merger events.}


\maketitle

\section{Introduction}
Planetary nebulae (PNe) are iconic astronomical objects, where the hot core of a former giant star
illuminates expelled envelope material. They do not only highlight spectacular mass loss episodes in stellar evolution, but they are also instrumental as standard candles in the cosmic distance ladder \citep{roth2021a}. In 80\%--85\% of cases, the nebulae are asymmetric with pronounced bipolar shapes and some show jet-like features \citep{sahai1998, parker2006}. 
Yet, the mechanism driving these bipolar, jet-like
outflows remains elusive.
The central stars of at least 15\% of PNe \citep{miszalski2009} are known to be close binary systems
but the true binary fraction is much higher, probably exceeding $60\%$
\citep{demarco2015, boffin2019}. This suggests that at least a large
fraction of PNe originates from common-envelope interaction in binary
stellar systems, where a star spirals into the envelope of a giant and
eventually ejects it \citep{paczynski1976, nordhaus2006, demarco2009, boffin2019}. 
  
Bipolar PNe are often observed to consist of two main components
\citep{sahai1998}: a toroidal structure of slowly expanding gas
($\approx20\,\kms$) and a perpendicular, fast bipolar (sometimes jet-like)
outflow \citep[$\approx 70\,\kms$, in many cases exceeding $100\,\kms$, see][]{imai2002, tafoya2020, guerrero2020}. 
Magnetic fields have been inferred in
such bipolar outflows, e.g., from synchrotron emission \citep{perez-sanchez2013},
dust continuum polarization \citep{sabin2015} and maser emission
\citep{miranda2001}, and are thought to help collimate them \citep{vlemmings2006}.
The low-velocity component has been associated with the common envelope (CE) of a
binary stellar system that is ejected preferentially into the region around its
orbital plane. Its approximate rotational symmetry with a polar cavity forms a
nozzle for a later fast outflow that is driven by an as yet unknown mechanism.
Ad-hoc central engines have been applied in toy models of post-CE systems and demonstrate that such models can indeed reproduce a wide variety of features found in PNe. Suggested engines include enhanced radiation-driven winds
\citep[spherically symmetric as well as bipolar;][]{morris1981, garcia-segura2005, zou2020}, 
magnetically-driven winds \citep{garcia-segura2020a} and jets launched from a
(circumbinary) disk \citep{soker1994, garcia-segura2021}.

Here, we explore the mechanism of forming bipolar PNe in CE interaction of binary systems in detailed and self-consistent magnetohydrodynamic (MHD) simulations. We briefly describe our simulations in Sect.~\ref{sec:methods}, present our main findings in Sect.~\ref{sec:results} and conclude in Sect.~\ref{sec:conclusions}.

\begin{figure*}
    \centering
	\includegraphics[width=\textwidth]{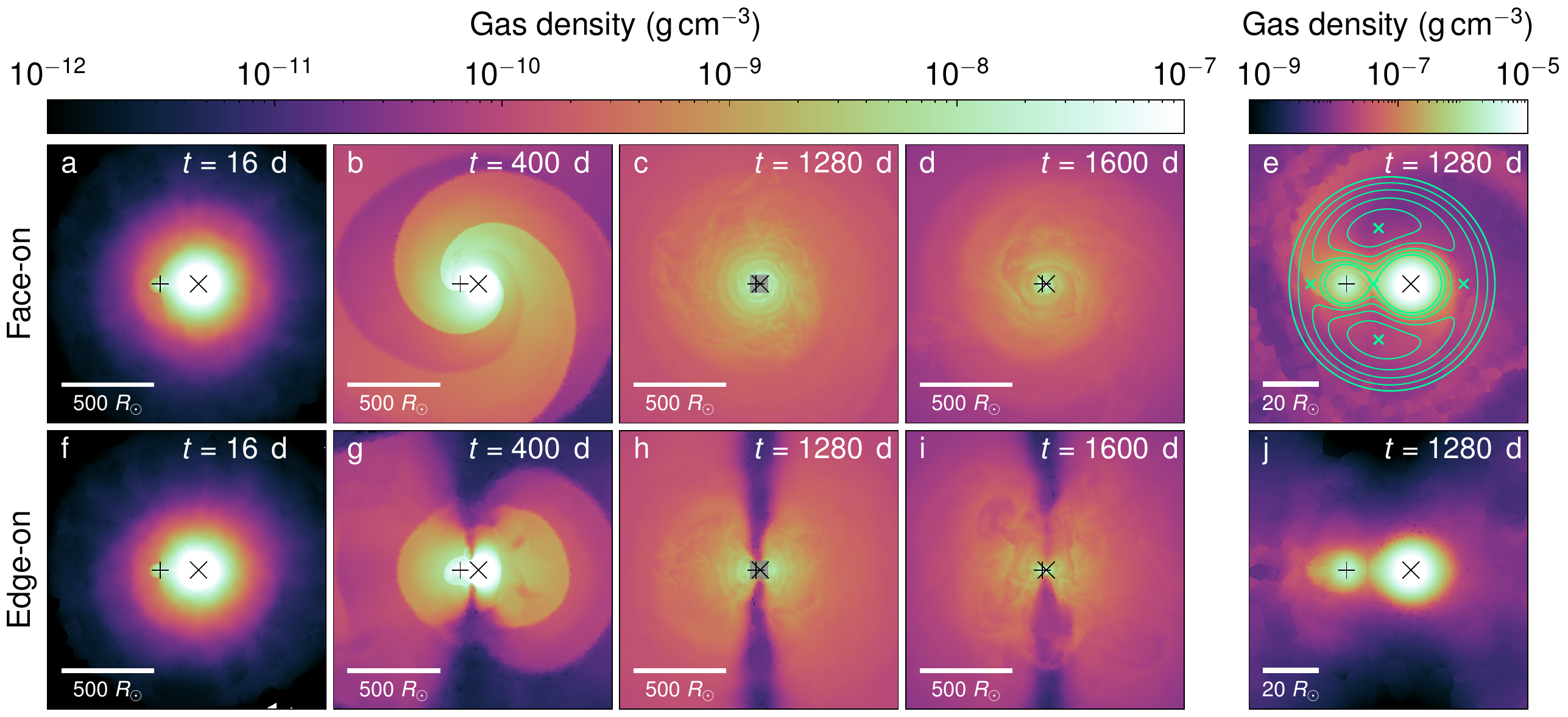}
	\caption{Density evolution during the common-envelope phase in our reference simulation. The panels are slices of gas density in the face-on view (top row) and edge-on view (bottom row). The cross symbol indicates the position of the core of the giant star and the plus symbol that of the companion. Panels (e) and (j) are zooms into the central region indicated by grey boxes in panels (c) and (h), respectively. The actual simulation domain is much larger than the panels shown here. Roche equipotential lines and the positions of the Lagrange points are added in panel (e).}
\label{fig:rho}
\end{figure*}

\section{Methods}\label{sec:methods}

Our simulations were performed in three steps, closely following earlier work \citep{ohlmann2016a, ohlmann2016b, kramer2020a, sand2020a}. First, we evolved a $1.2 \,
\msun$ zero-age main sequence star at solar metallicity with the
one-dimensional stellar evolution code \mesa
\citep{paxton2011a,paxton2013a,paxton2015a} in version 7624. We used the default \mesa settings and accounted for
stellar wind mass loss via the  
\citet{reimers1975a} prescription with wind scaling factor $\eta = 0.5$
for red giant stars and via the \citet{bloecker1995a} prescription with $\eta = 0.1$ for
asymptotic-giant-branch (AGB) stars. For the CE simulations presented here, 
we constructed an early AGB model that has a mass of $\mprim = 0.970 \, \msun$ and a radius of $173\,
\rsun$ by the time CE interaction commences. Following the procedure of \citet{ohlmann2017a}, we replace the $0.545\, \msun$ core of this stellar model by a point particle that interacts only
gravitationally with the gas of the envelope.

Second, we mapped the primary star into the magnetohydrodynamic moving-mesh code \arepo \citep{springel2010, pakmor2011, pakmor2013} and
performed a relaxation as described by \citet{ohlmann2017a} to re-establish
hydrostatic equilibrium in the envelope. Both, the relaxation step and the subsequent CE simulations use the OPAL equation of state \citep{rogers1996a, rogers2002a}. The overall relaxation was carried out for ten dynamical timescales of the primary star with half of the time damping spurious velocities and the other half ensuring a hydrostatically stable model. At the end of the relaxation, the star is resolved by about 5 million cells.

Third, we added a $\mcomp = 0.243\, \msun$ companion, that could be a white-dwarf or a main-sequence star, at a distance of $207 \, \rsun$ from the center
of the primary star. The mass ratio between the companion and the primary star in this \emph{reference simulation} is $q = 0.25$. To explore the effect of varying mass ratios, two additional setups were used with $q=0.5$ ($\mcomp = 0.486\, \msun$, placed at a distance of $236\, \rsun$) and $q=0.75$ ($\mcomp = 0.729\, \msun$, initial distance $256 \, \rsun$). In all our simulations, the companion is unresolved and represented by
a point particle in the same way as the core of the primary star. The
stellar envelope was initially set to solid-body rotation at $95\%$ of the orbital velocity
of the binary system. We employ adaptive mesh refinement to ensure high resolution around both point particles.

Our setup parameters are the same as those of \citet{sand2020a}, except for a seed magnetic dipole field that we added to the primary star's envelope in our reference simulation labeled ``MHD''. As a comparison, one simulation (labeled ``non-MHD'') was performed with identical parameters but no initial magnetic field. The initial polar surface 
field strength was set to $10^{-10} \; \gauss$, similar to the models of \citet{ohlmann2016b}. Using the OPAL equation of
state in all our simulations, we followed changes in the
ionization state of the gas caused by the expansion of the stellar envelope. In our
model, released recombination energy is assumed to thermalize locally. If converted into kinetic energy, it may support envelope ejection.

\section{Results}\label{sec:results}

\subsection{Dynamical evolution during the common envelope phase}

After the companion is placed at a distance of $207 \, \rsun$ from the center of
the primary star (Figs.~\ref{fig:rho}a and
f), it is immediately dragged
into the envelope of the giant (see also videos in Figs.~\ref{fig:mov-rho-face} and \ref{fig:mov-rho-edge}). 
The core of the
primary star and the companion orbit each other inside a common envelope causing
a spiral density structure (Fig.~\ref{fig:rho}b).

\begin{figure}
	\centering
	\resizebox{\hsize}{!}{\includegraphics{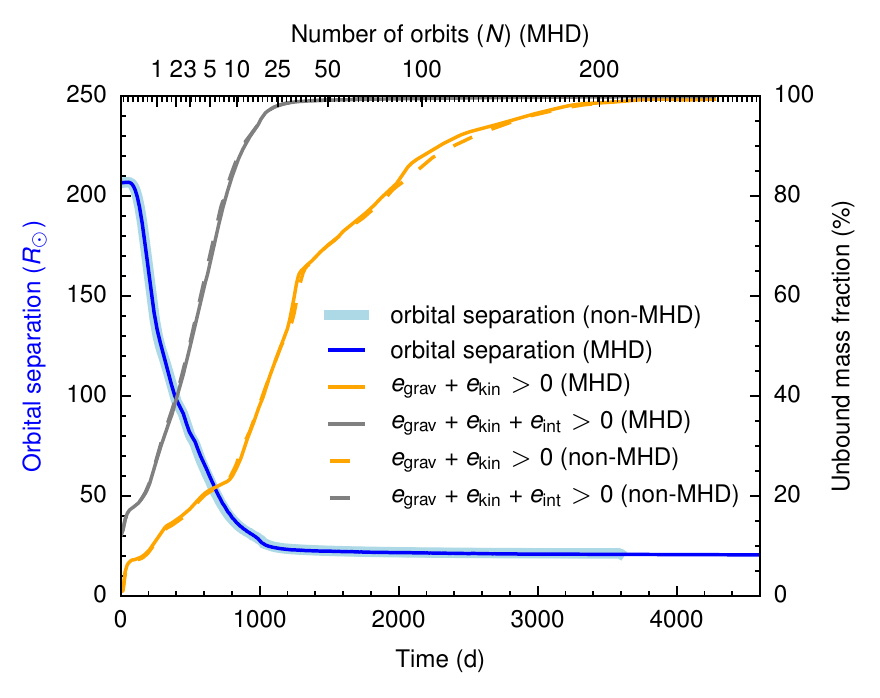}}
	\caption{Evolution of orbital separation and unbound mass fraction of the envelope in the simulations taking into account magnetic fields (MHD) and neglecting magnetic fields (non-MHD). In both simulations, (MHD and non-MHD), local thermalization of released recombination energy is assumed. For the non-MHD simulation we show only data until $t=3568 \, \dday$. The unbound mass fraction is computed according to a kinetic-energy criterion (orange) and an internal-energy criterion which additionally includes ionization energy released in recombination processes (gray).}
	\label{fig:orbit-and-unbound-mass}
\end{figure}

Drag forces on the companion and on the core of the giant transfer angular momentum and energy to the envelope gas, leading to orbital decay. In Fig.~\ref{fig:orbit-and-unbound-mass}, we compare the orbital separation between the core of the AGB star and the companion in our reference MHD case and its corresponding non-MHD version. In both simulations, the orbital separations decrease similarly from $207 \, \rsun$ to $21.5 \, \rsun$ (non-MHD) and $21.0 \, \rsun$ (MHD). The final orbital periods of the systems are $13.1 \, \dday$ (non-MHD) and $12.6 \, \dday$ (MHD). By the time the MHD simulation is stopped, only $1.65 \times 10^{-4} \msun$ of gas remains in the vicinity of the cores. The generated magnetic-field energy ($E_{\mathrm{mag}}\approx 10^{43} \, \mathrm{erg}$, see Sect.~\ref{sect:bamp}) is smaller by a factor of $1000$ than the kinetic and potential energies of the cores and the envelope ($\approx 10^{46} \, \mathrm{erg}$). Hence, magnetic fields do not influence the final orbital separation of the system significantly. The dynamical plunge-in ends when
the orbital decay slows down at ${\sim}\, 1000 \, \dday$ (${\sim}\, 25$ orbits; see Fig.~\ref{fig:orbit-and-unbound-mass}), and the spiral structure is washed out by the many orbits of the remnant
binary around the common center of mass (Fig.~\ref{fig:rho}c).

We use two criteria to measure the fraction of unbound to total envelope mass: A lower limit to the unbound mass is given by the \emph{kinetic-energy criterion,} that includes only the kinetic, $e_{\mathrm{kin}}$, and the potential energy, $e_{\mathrm{grav}}$, of each gas cell (orange lines in Fig.~\ref{fig:orbit-and-unbound-mass}). According to this criterion, mass is unbound if $e_{\mathrm{grav}} + e_{\mathrm{kin}} > 0$. The \emph{internal-energy criterion,} $e_{\mathrm{grav}} + e_{\mathrm{kin}}+ e_{\mathrm{int}} > 0$ (gray lines in Fig.~\ref{fig:orbit-and-unbound-mass}), additionally includes the thermal and ionization energy of the gas cells, $e_{\mathrm{int}}$, and provides an upper limit. The contribution of magnetic field energy is negligible. There is almost no difference in the unbound mass fractions as given by the internal-energy criterion between the MHD and the non-MHD simulations.
Applying the kinetic-energy criterion, we find a slightly faster mass unbinding in the MHD simulation. Nevertheless, the unbound mass fractions converge to very similar values in both cases. Almost the entire envelope is unbound as measured by the kinetic-energy criterion (99.4 \% MHD and 98.8\% non-MHD). The success of almost complete envelope ejection is therefore independent of magnetic fields and is mainly due to including recombination energy in the simulations. 

Here, the released recombination energy is assumed to thermalize locally. This appears to be a reasonable assumption because most of the recombination of hydrogen takes place inside a volume enclosed by the photosphere (see Fig.~\ref{fig:photosphere} in Appendix~\ref{sec:photosphere}). In addition, helium recombines close to the center of the envelope in optically thick regions (Fig.~\ref{fig:photosphere}). While this justifies our assumption of local thermalization of recombination energy within our simulation it does not prove that photon diffusion and radiative losses are completely negligible. Our models may be viewed as an upper limit to envelope ejection. \cite{sand2020a} present a simulation for the same setup without magnetic fields that neglects ionization energy in the equation of state. They then indeed find that only about 16\% of the envelope is ejected, which can be considered as a lower limit. Given that most of the recombination is found to occur inside the optically thick part of the envelope, we consider almost full envelope ejection the more likely outcome for the binary system studied here. Ultimately, however, this has to be confirmed with models that account for potential losses of recombination energy by radiation.

In order to compare the binding energy of the envelope of the primary star, $E_{\mathrm{bin}}$, to the orbital energy released during the inspiral process, $\Delta E_{\mathrm{orb}}$, we compute the CE ejection efficiency
$\alpha_{\mathrm{CE}} = E_{\mathrm{bin}} / \Delta E_{\mathrm{orb}}$. Here, $E_{\mathrm{bin}}$ and $\Delta E_{\mathrm{orb}}$ are defined as in \cite{sand2020a}, and we consider the two cases that (i) only gravitational energy ($\alpha_{\mathrm{CE,g}}$), and (ii) gravitational plus internal energy ($\alpha_{\mathrm{CE,b}}$) are taken into account in the computation of $E_{\mathrm{bin}}$. In our reference simulation we obtain $\alpha_\mathrm{CE,g} = 2.87$ and $\alpha_\mathrm{CE,b} = 0.32$ at the time when the calculation is terminated. Our values are in good agreement with those found in \cite{sand2020a} and differ only slightly because our new MHD simulations are evolved longer until almost full envelope ejection is reached.

During the plunge-in of the companion, the now centrifugally-supported envelope starts to expand into a
toroidal structure in the orbital plane with low
density funnels around the polar axis (Fig.~\ref{fig:rho}b and g). While the envelope expands, gas with low specific angular momentum falls back onto the central binary along the
orbital angular-momentum vector. This transforms the funnels into a narrow, low-density chimney (Fig.~\ref{fig:rho}h).

\begin{figure}
	\centering
    \resizebox{\hsize}{!}{\includegraphics{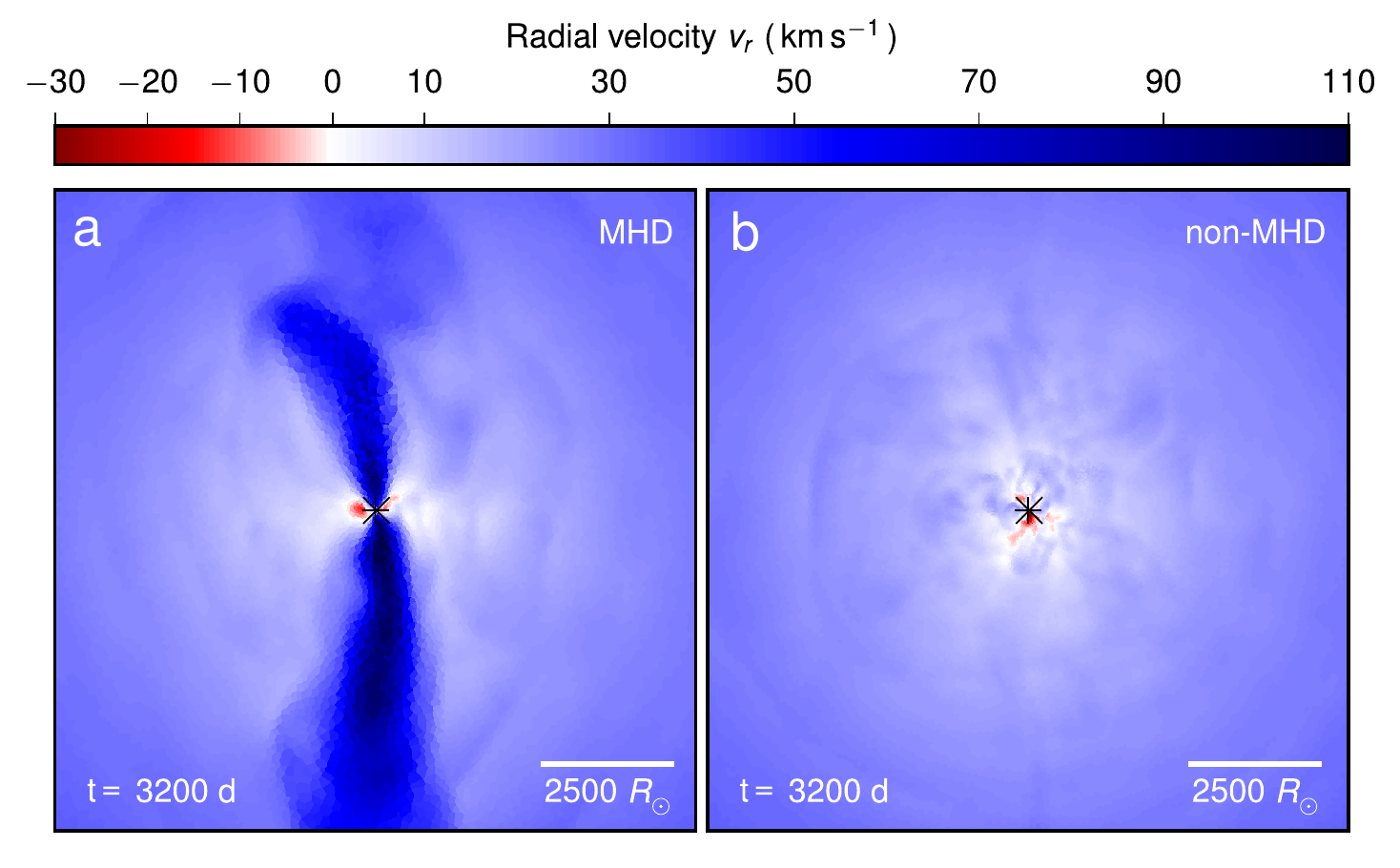}}
    	\caption{Radial velocity in the MHD reference simulation (a) and the non-MHD simulation (b) plotted edge-on.}
	\label{fig:mhd-nonmhd}
\end{figure}

\begin{figure*}
    \centering
	\includegraphics[width=17cm]{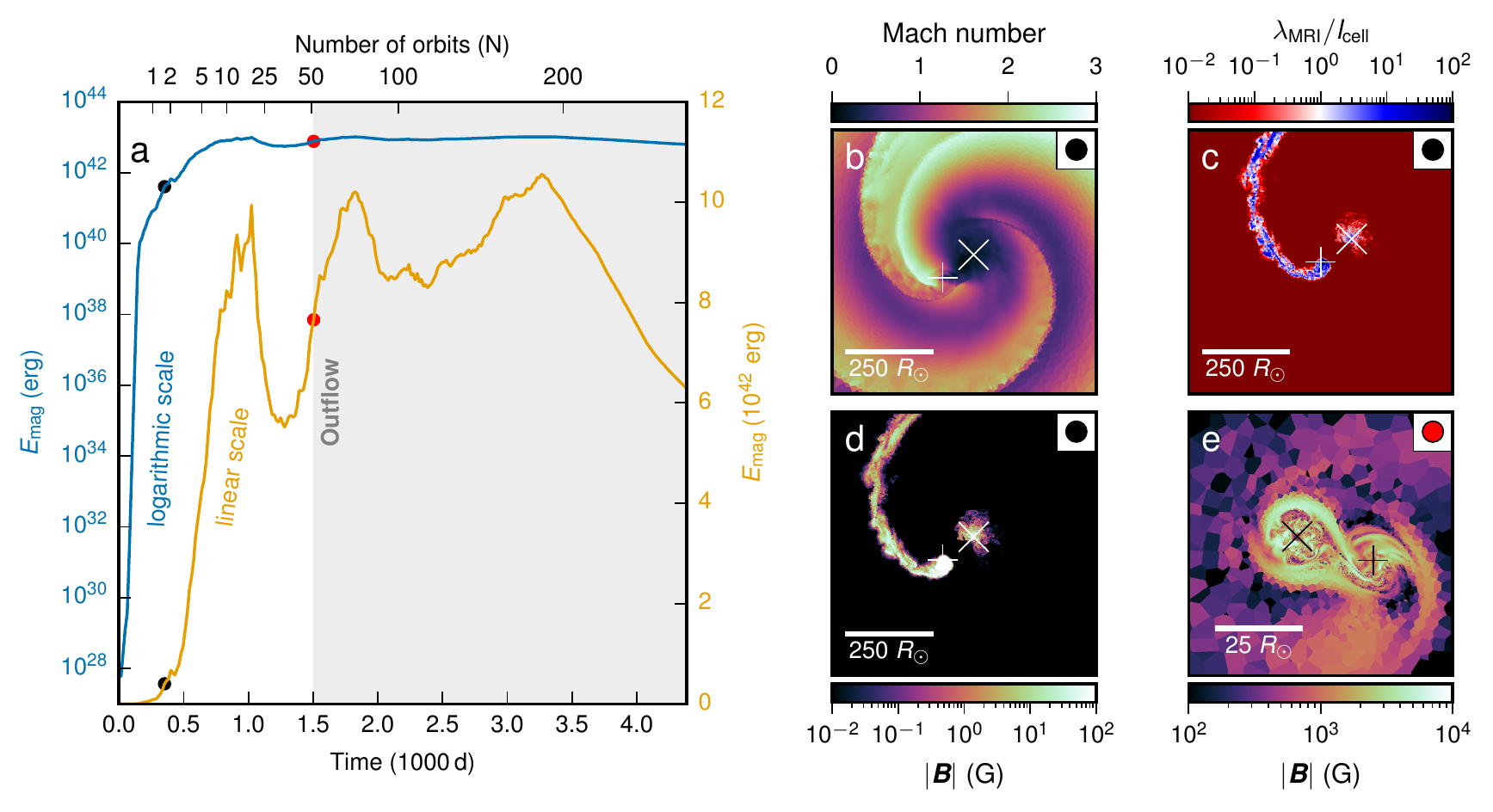}
	\caption{Amplification of magnetic fields. In panel (a), the total magnetic energy is shown as a function of time on a logarithmic scale (blue) and on a linear scale (orange). Panels (b), (c), (d) and (e) show the sonic Mach number, the ratio of the wavelength of the fastest growing MRI mode and the \arepo cell size $\lambda_\mathrm{MRI}/l_\mathrm{cell}$, and the absolute magnetic field strength $|\vec{B}|$ in the orbital plane, respectively. They provide details
	of the two episodes of field amplification at $t = 352 \, \mathrm{d}$, marked with black dots, and at $t = 1504 \, \mathrm{d}$, indicated with red dots.}
	\label{fig:eb}
\end{figure*}

\begin{figure*}
    \centering
    \includegraphics[width=16cm]{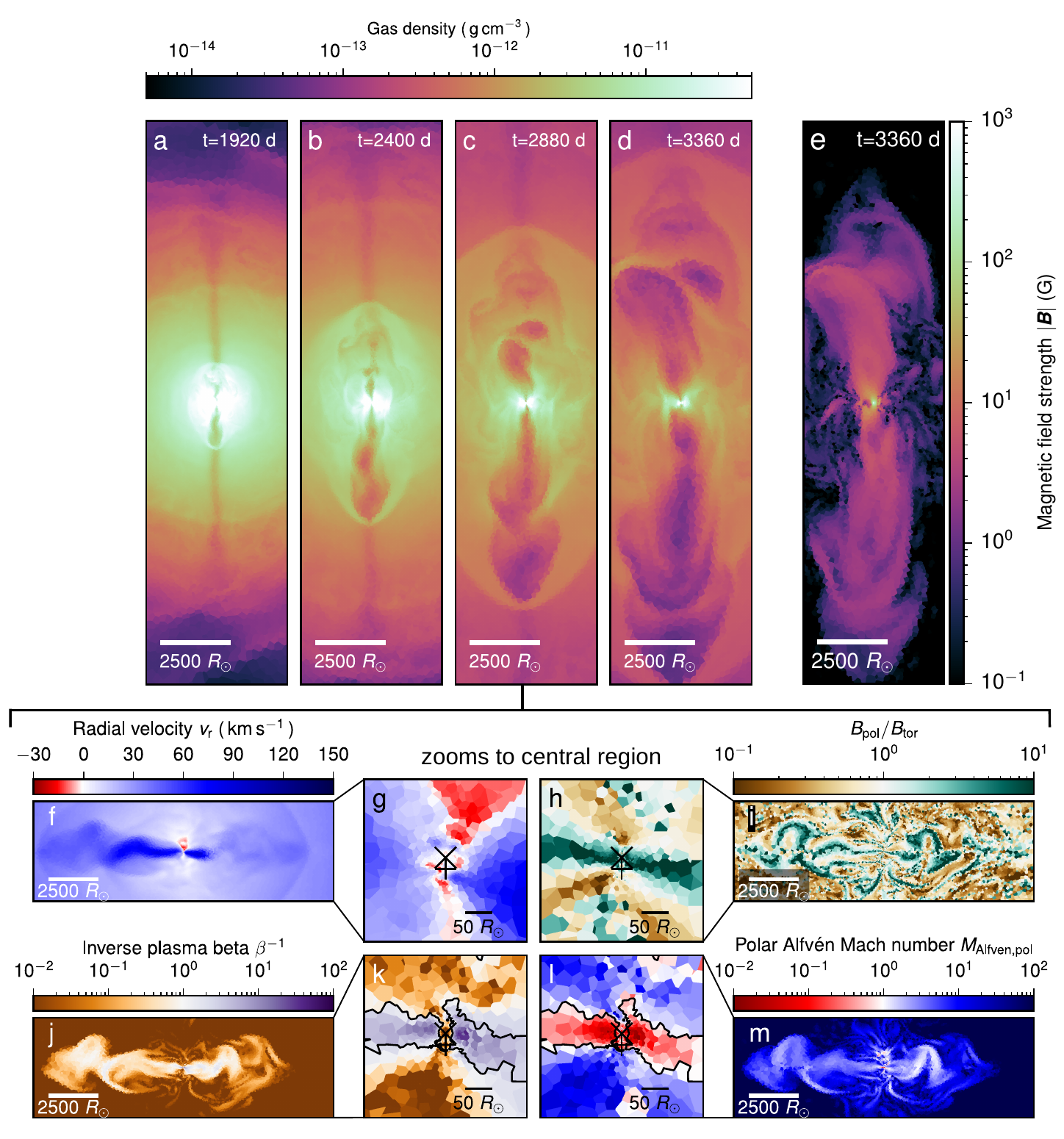}
	\caption{Evolution and launching of the bipolar outflow perpendicular to the orbital plane. The gas density in panels (a)--(d) traces the evolution of the outflow, and in panel (e) we show the absolute magnetic field at the same time as in panel (d). The bottom panels (f)--(m) (rotated by $90^\circ$ with respect to the upper panels) are at a time of $2880\,\mathrm{d}$ and illustrate the launching of the outflow. Shown are radial velocity $v_\mathrm{rad}$ (f, g), ratio of poloidal to toroidal magnetic field strength $B_\mathrm{pol}/B_\mathrm{tor}$ (i, h), ratio of magnetic to thermal pressure $\beta^{-1}$ (j, k) and poloidal \alfven Mach number $M_{\mathrm{alfven,pol}} = v_{\mathrm{pol}} \sqrt{4\pi \rho} / B_{\mathrm{pol}}$. Panels (g), (h), (k) and (l) are zooms into the central region around the binary. The black contours in panels (k) and (l) indicate ``force-free'' regions inside which both $\beta^{-1} > 1$ and $M_{\mathrm{alfven,pol}} < 1$.}
	\label{fig:outflow}	
\end{figure*}

Shortly after $1000\,\mathrm{d}$, the envelope is virtually unbound according the the internal-energy criterion.
This, however, is not the end of the dynamical interaction. At about this time, a
contact phase sets in between the companion and the former core of the giant
star (Figs.~\ref{fig:rho}e and j) where both stars overfill their respective
Roche lobes as indicated by the Roche potential in Fig.~\ref{fig:rho}e. At
${\sim}\,1500\,\mathrm{d}$, we observe a new and important feature in our CE
simulation: a high-velocity outflow ($90-130 \, \kms$) emerges
perpendicular to the orbital plane from the region of the central core binary. This is visible in the density structure
(Fig.~\ref{fig:rho}i) but more clearly in the radial velocity component shown in Fig.~\ref{fig:mhd-nonmhd}a.
The outflow builds up gradually, is first
launched into one hemisphere only, and precesses.
Once fully established,
it becomes bipolar and is collimated along the
chimney-like underdensity, embedded in the remainder of the common envelope. By
the time it launches, at least 70\% of the envelope of the former giant star has
been ejected. Our non-MHD comparison simulation undergoes a nearly identical evolution, resulting in a similar envelope mass ejections but it develops no fast polar
outflow. Instead, it shows an almost spherical expansion of the envelope with only minor angular variations of the velocity (Fig.~\ref{fig:mhd-nonmhd}b). Together with the fact that the magnetic
pressure in the outflowing material is larger than the thermal pressure
(Fig.~\ref{fig:outflow}j), this implies that the outflow is magnetically driven. 
It is thus an inevitable consequence of the magnetic field amplification 
that we observe in our CE simulation.

\subsection{Amplification of magnetic fields}
\label{sect:bamp}

During the plunge-in, the total magnetic energy in the envelope increases
exponentially by more than 15 orders of magnitude (Fig.~\ref{fig:eb}a). The
strongest amplification of the magnetic field is observed in the region around
the companion star (Fig.~\ref{fig:eb}d), which accretes about $10^{-5} \, \msun$ while moving
through the envelope, and in a shear layer between the spiral arm emerging at the
companion and the rest of the envelope. We identify three mechanisms of magnetic
field amplification:
\begin{enumerate}[(i)]
\item The magneto-rotational instability
\citep[MRI,][]{balbus1991}, which acts close to the companion and near the core of the giant star. 
A comparison of the wavelength of the fastest-growing mode to the size of
the \arepo grid cells demonstrates that the MRI is well resolved in these
regions (Fig.~\ref{fig:eb}c).
The $e$-folding growth time of the magnetic energy in our simulation is about $8\,\mathrm{d}$, which is on the order of the expected MRI growth time\footnote{The MRI wavelength and growth time are computed as in \citet{rembiasz2016a}} of ${\approx}\,15$--$20\,\mathrm{d}$.
\item Kelvin-Helmholtz
instabilities, which occur in shear layers that are visible in Mach number (Fig.~\ref{fig:eb}b) in combination with the MRI. The amplified field is seen in the region trailing the companion (``$+$'' symbol, see Fig.~\ref{fig:eb}d).
\item Convective fluid motions as in dynamo processes, which create an isentropic region around the core of the primary (Fig.~\ref{fig:eb}d, region around the ``$\times$'' symbol).
\end{enumerate}

The MRI amplifies the magnetic field close to the positions of the core of the primary and the companion. Within a sphere of radius $r=8.7 \,\rsun$ around the point particles, the gravitational potential is softened using a spline function. 
The artificially softened gravitational potential reduces the gradients of the flow velocities, which tends to underestimate the amplification processes.

After the initial fast orbital decay, at ${\sim}\, 1000\, \mathrm{d}$, 
the magnetic field energy reaches a maximum of about
$10^{43}\,\mathrm{erg}$. This terminates the first phase of magnetic field amplification (Fig.~\ref{fig:eb}a). The
magnetic flux is conserved in ideal magneto-hydrodynamics, which causes
a decrease of the magnetic field energy when the envelope starts to expand
significantly (Fig.~\ref{fig:eb}a, between $1000 \, \dday - 1300 \, \dday$). This
decrease is reversed at ${\gtrsim}\, 1300\,\dday$, i.e.\ at about the time when the
magnetically driven outflow emerges, because gas flows that establish when the core binary enters a contact phase lead to further magnetic field amplification (Fig.~\ref{fig:eb}e). 
This amplification is driven by shear layers in the gas streams subject to the MRI, similar to the situation in the merger of two main-sequence stars \citep{schneider2019}.
In the amplification region, the wavelength of the fastest growing MRI mode is larger by a factor 
of 10 to 100 than the typical cell size. 
In the Roche-lobe-overflow material, the magnetic field energy grows to a similar order of magnitude as the total kinetic energy of the gas leading to equipartition values of $E_{\mathrm{mag}} / E_{\mathrm{kin}} \approx 1$. 
In some regions even superequipartition is reached. This is known from fast-rotating systems \citep[e.g.][]{augustson2016a} and indicates dynamo action in conjunction with the MRI. 
The mass exchange in the core binary is non-conservative and highly magnetized material is lost through the outer Lagrangian points. 
This continued magnetic field amplification in the contact binary eventually launches the high-velocity 
outflow parallel to its axis of rotation (see video in Fig.~\ref{fig:mov-jet}).

\subsection{Launching of the outflow}
\label{sect:outflow}

The complex flows close to the core binary system (e.g.\
Fig.~\ref{fig:rho}e, j) do not reproduce the idealized picture of an
accretion or circumbinary disk often associated with the launching of
magnetized outflows. Still, the global symmetry and the motion of the stellar
cores in the orbital plane create a strong, preferentially poloidal magnetic
field along the chimney that has a diameter of about the binary's orbital
separation (Fig.~\ref{fig:outflow}h). The outflowing material is
accelerated to velocities of $90-130 \, \kms$ (Fig.~\ref{fig:outflow}f)
approximately matching the ${\sim}\,85 \, \kms$ orbital velocity of the core binary. The material in the outflow corotates at a lower speed of about ${\sim}\,40 \, \kms$.

The flow structure in the acceleration region is characterized by a small poloidal
\alfven Mach number $M_\mathrm{A}= 10^{-2}<1$ (Fig.~\ref{fig:outflow}l).
This is a typical signature of a magneto-centrifugally driven outflow
launched by the Blandford--Payne mechanism \citep{blandford1982}. In Figs.~\ref{fig:outflow}k and~\ref{fig:outflow}l, we indicate the ``force-free'' regions where both $\beta^{-1} = P_{\mathrm{mag}} / P_{\mathrm{gas}} > 1$ and $M_{\mathrm{A}} < 1$. In these regions, the magnetic pressure exceeds the thermal and the dynamic pressure, and fluid elements can therefore be accelerated along the magnetic field lines by centrifugal forces.
Additionally, a magnetic pressure gradient of the poloidal field contributes to the acceleration of material. 

In the circumbinary envelope material,
however, the magnetic field is dominated by its toroidal component, which results from strong differential rotation (Fig.~\ref{fig:outflow}i). 
It contributes to the overall acceleration of the
bipolar outflow -- an effect similar to magnetic tower jets \citep{uchida1985, lyndenbell1994} that, \eg, arise during star formation \citep[e.g.][]{pudritz2019}. 
The outflow is replenished by
accretion of left-over envelope material in the orbital plane as indicated by the
negative radial velocity in Fig.~\ref{fig:outflow}g.
As the high-velocity outflow propagates along the collimating chimney inside a
higher-density cocoon (Figs.~\ref{fig:outflow}b to d), its interaction with the
expanding envelope material creates bow shocks with sonic Mach numbers of 2--3.

\subsection{Sensitivity to mass ratio}

\begin{figure}
	\centering
	\resizebox{\hsize}{!}{\includegraphics{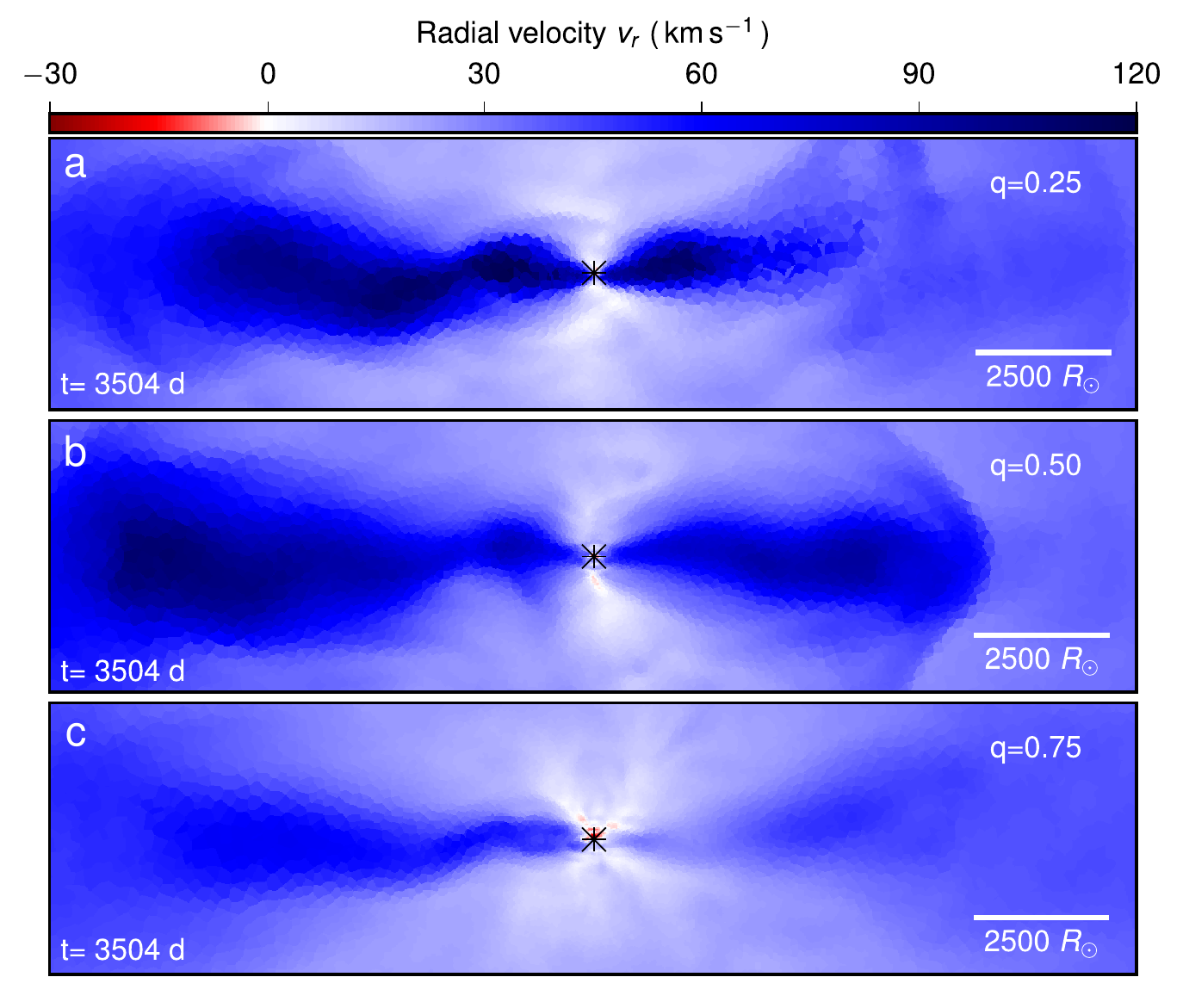}}
	\caption{Radial velocities in simulations with different mass ratios $q$ of companion and primary-star mass as indicated.}
	\label{fig:parameter}
\end{figure}

To explore the universality of the mechanism producing high-velocity outflows, 
we have carried out two
additional MHD simulations with initial mass ratios of $q=0.5$ and $q=0.75$. In the case of $q=0.50$, the orbit shrinks to a final separation of $37.8 \, \rsun$, corresponding to a period of $26.9 \, \dday$. For $q=0.75$, the remaining stars are separated by $54.2 \, \rsun$  (orbital period: $65.8 \, \dday$) by the time the computation is stopped. The distance between the two stars is still decreasing slightly at the end of all simulations. During the simulated time, the model with mass ratio $q=0.50$ ($q=0.75$) unbinds at least 99.8\%  (97.0\%) of the gas mass according to the kinetic-energy criterion.

As can be seen in Fig.~\ref{fig:parameter}, pronounced bipolar outflows are launched in all simulations regardless of the initial mass ratio of the binary. We observe different maximum outflow velocities. In the models with mass ratios $q=0.25$ and $q=0.50$, radial velocities of ${>}\, 120\, \kms$ dominate the outflow whereas in the simulation with the largest companion mass ($q=0.75$) the fastest velocities are on the order of $60-70 \, \kms$. It is possible that the strength of the outflow in the latter simulation will further increase. However, as discussed in Sect.~\ref{sect:outflow}, the outflow velocities are related to the orbital velocities of the remnant binary. Given the tighter orbits reached in simulations with lower-mass companions, we indeed expect faster outflows in such cases as indicated in Fig.~\ref{fig:parameter}.

\subsection{Comparison to observations}

\begin{figure}
    \centering
	\resizebox{\hsize}{!}{\includegraphics{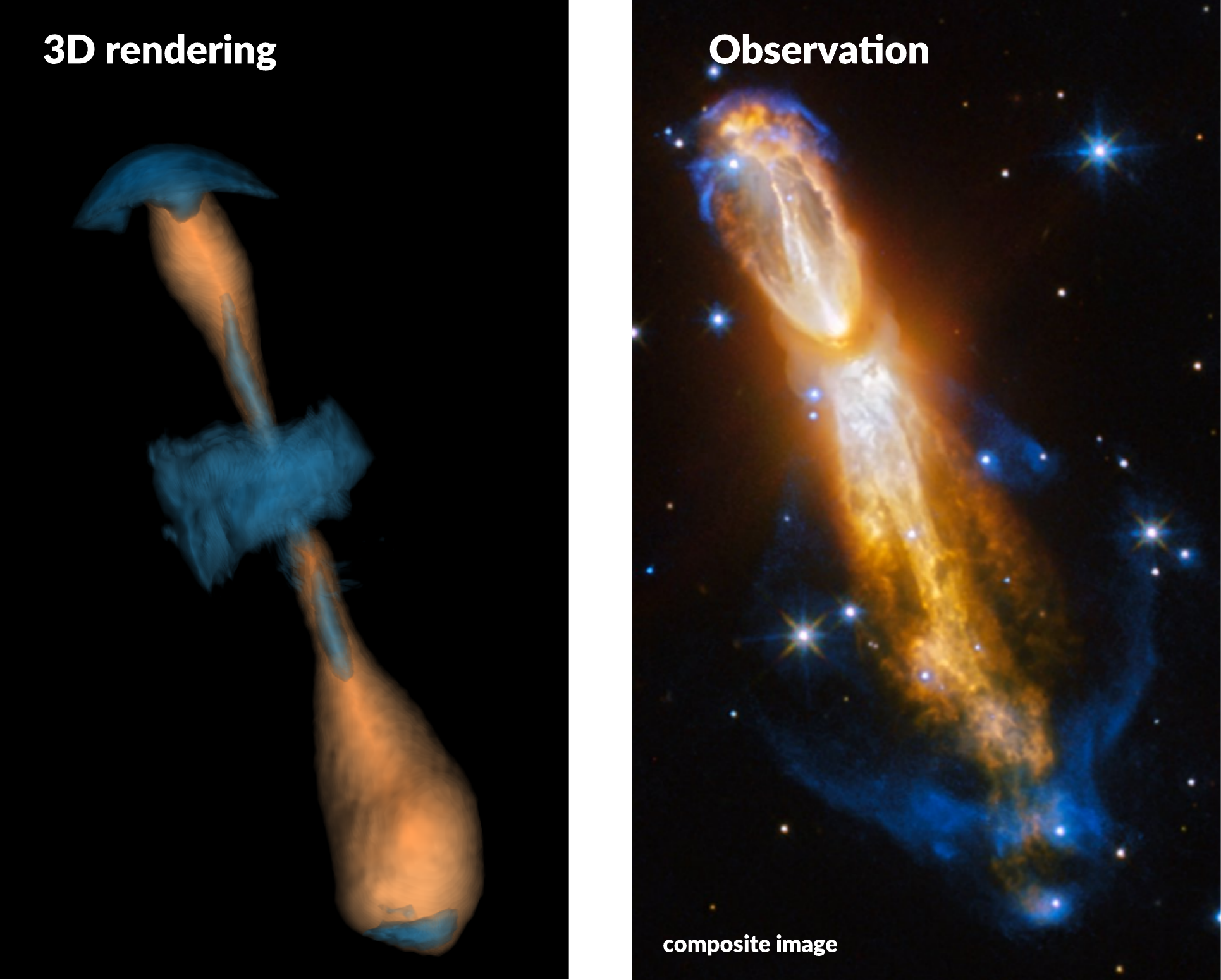}}
	\caption{Qualitative morphological comparison between the simulated outflow and observations of the Calabash nebula (OH~231.84~+4.22). Left panel: geometry of the outflow represented by radial velocity (orange) and temperature (blue). Right panel: composite, optical image taken by the NASA/ESA Hubble Space Telescope. Note that the rendering of our simulation is not to scale with the observation.}
	\label{fig:planetary-nebula}
\end{figure}

The bipolar outflow and the structure of the ejected envelope define the
morphology of the forming planetary
nebula. To compare the obtained outflow structure with observations,
we show a 3D rendering of our model at $4400\,\mathrm{d}$ and a Hubble Space Telescope image of the protoplanetary Calabash
Nebula (OH 231.8+04.2) in Fig.~\ref{fig:planetary-nebula}. Our model is neither meant to reproduce this particular
nebula quantitatively nor does it predict the observables directly, but the morphological similarities are striking. A lopsided bipolar, jet-like
outflow is launched from a central toroidal structure inside which a binary
central star is present. It terminates in bow shocks on both ends visible as high temperatures at the top and bottom of the bipolar outflow (Fig.~\ref{fig:planetary-nebula}). The high temperature region at the waist of the structure originates from the circumbinary material around the central contact binary.

The
bipolar outflow of the Calabash nebula is known to be threaded by an
ordered magnetic field \citep{leal-ferreira2012, sabin2015}, which further supports our model (Fig.~\ref{fig:outflow}e). The expansion
velocity of our toroidal structure of order $10\text{--}30\,\kms$ (Fig.~\ref{fig:outflow}f)
and the $90-130 \, \kms$ bipolar outflow are broadly consistent with
observations of young (proto)PNe and post-AGB stars \citep{imai2002, tafoya2020, guerrero2020}. In our simulation, we measure mass outflow rates of $\dot{M} \simeq 10^{-5} - 10^{-3} \, \msun \,\mathrm{yr}^{-1}$, which is consistent with observational findings \citep{lorenzo2021a}.
This shows that
our proposed mechanism self-consistently reproduces the structures observed in
such systems.

\section{Discussion and conclusions}\label{sec:conclusions}

In analogy to the robustness against variations in the mass ratio of the system, we expect that the same mechanism operates in
common-envelope phases of primary stars with different masses. The morphology of the ejected common envelope material depends on the initial parameters of the system and some simulations show less pronounced chimney structures \citep{ohlmann2016a}, which offers an explanation for the variety of shapes observed in PNe. We therefore predict that magnetically driven
outflows launched by the core binary in late stages of the dynamical CE
evolution are responsible for a significant fraction of the observed PNe and
post-AGB stars. This is in agreement with results from population synthesis studies \citep{han1995}.

Orbital separations found in CE simulations are typically too large compared with observations of post-CE systems \citep{iaconi2017}. In principle, the bipolar outflows found in this study could help alleviate this issue if they persist for longer times after the dynamical CE phase and carry away further angular momentum from the remnant binary. Given the low mass of bound envelope gas remaining in the system by the end of our simulation, however, we expect little impact on the final orbital separation in the case studied here.

Our simulations establish a direct link between some bipolar PNe and the enigmatic common-envelope phase of binary star evolution. PNe may thus be promising laboratories to gain further insights into this evolutionary phase that is pivotal also for the formation of X-ray binaries, Type Ia supernovae and gravitational-wave sources
\citep{ivanova2013}.

\begin{acknowledgements}
We thank Giovanni Leidi for discussions about magnetic field amplification processes and the reviewer for useful comments that helped improve our work.
This work has received funding from the European Research Council (ERC) under the European Union’s Horizon 2020 research and innovation programme (Grant agreement No.\ 945806). PAO, FKR, FRNS, and CS acknowledge funding from the Klaus Tschira foundation. This work is supported by the Deutsche Forschungsgemeinschaft (DFG, German Research Foundation) under Germany’s Excellence Strategy EXC 2181/1-390900948 (the Heidelberg STRUCTURES Excellence Cluster).
\end{acknowledgements}

\bibliographystyle{bibtex/aa} 

\begin{appendix} 

\section{Movies of the common-envelope simulation}\label{sec:movies}

\begin{figure}
	\centering
	\resizebox{\hsize}{!}{\includegraphics{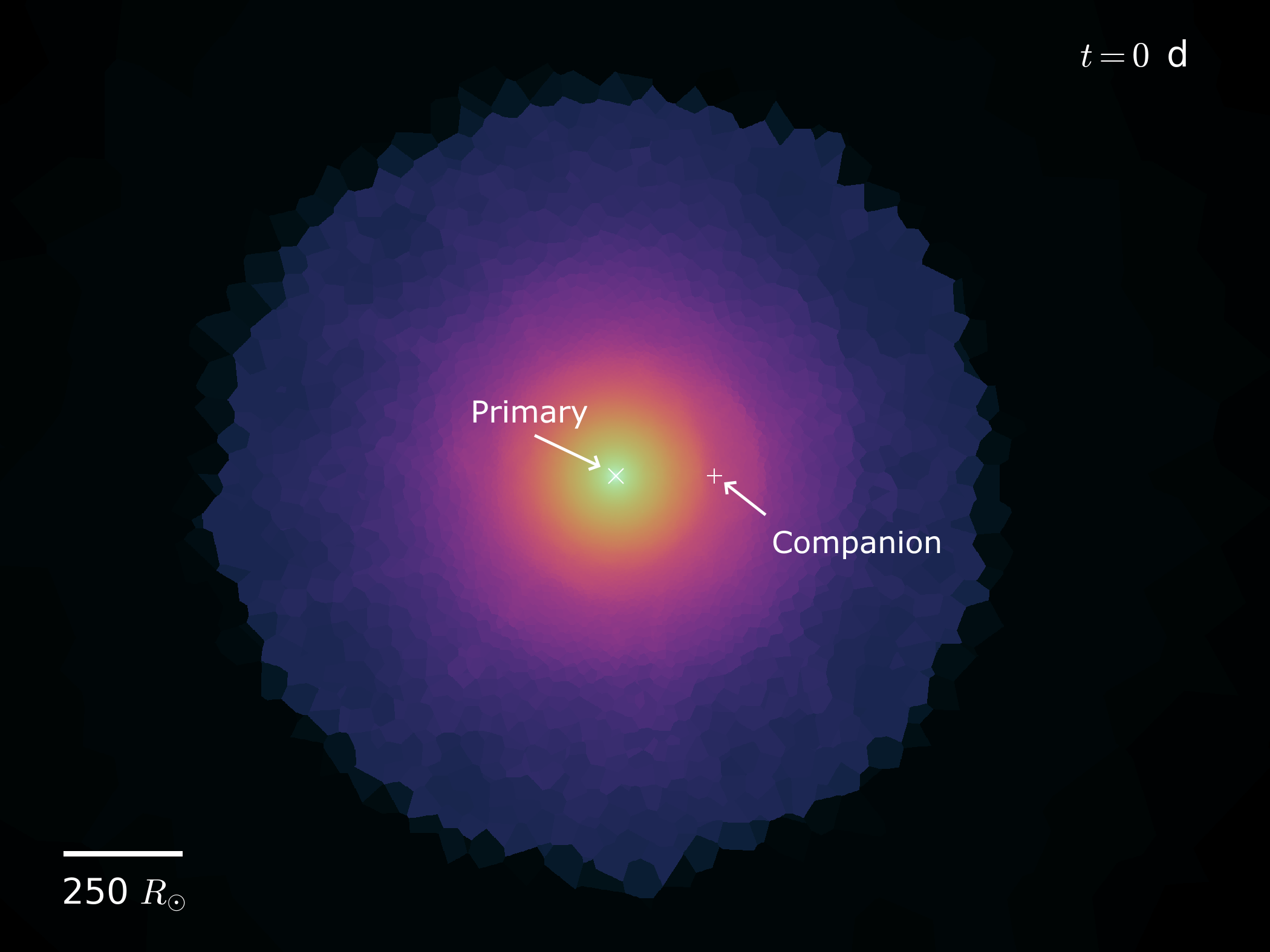}}
	\caption{(\href{https://zenodo.org/record/5596192/files/fig_A1_movie_density_face_on.mp4?download=1}{Movie online}) Movie of density evolution in the orbital plane.
Similar to Fig.~\ref{fig:rho}, in this animation the evolution of the density is shown in the orbital plane.}
	\label{fig:mov-rho-face}
\end{figure}

\begin{figure}
	\centering
	\resizebox{\hsize}{!}{\includegraphics{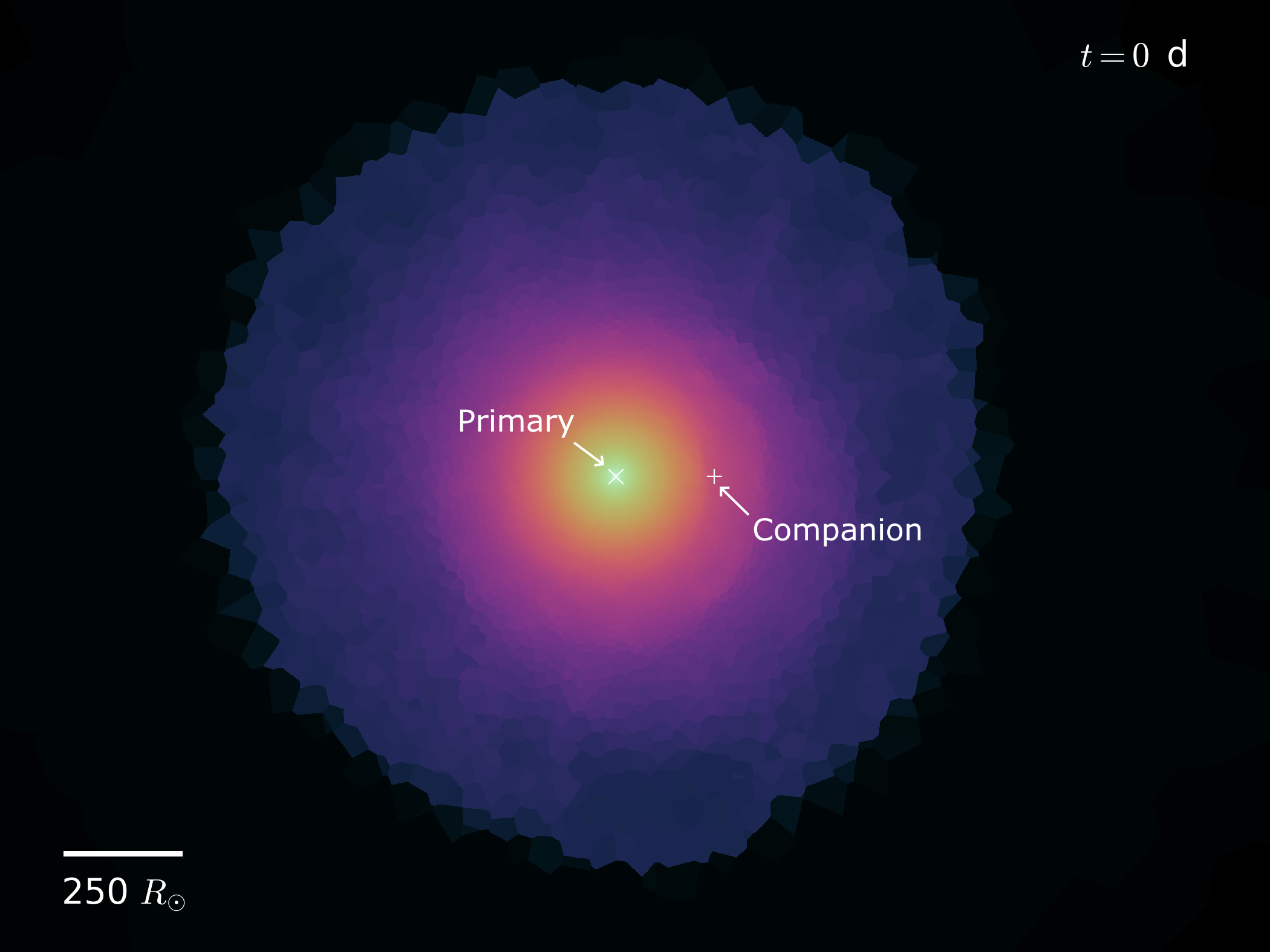}}
	\caption{(\href{https://zenodo.org/record/5596212/files/fig_A2_movie_density_edge_on.mp4?download=1}{Movie online}) Movie of density evolution in the plane perpendicular to the orbital plane.
Similar to Fig.~\ref{fig:rho}, in this movie the evolution of the density is shown in the plane perpendicular to the orbital plane.}
	\label{fig:mov-rho-edge}
\end{figure}

\begin{figure}
	\centering
	\resizebox{\hsize}{!}{\includegraphics{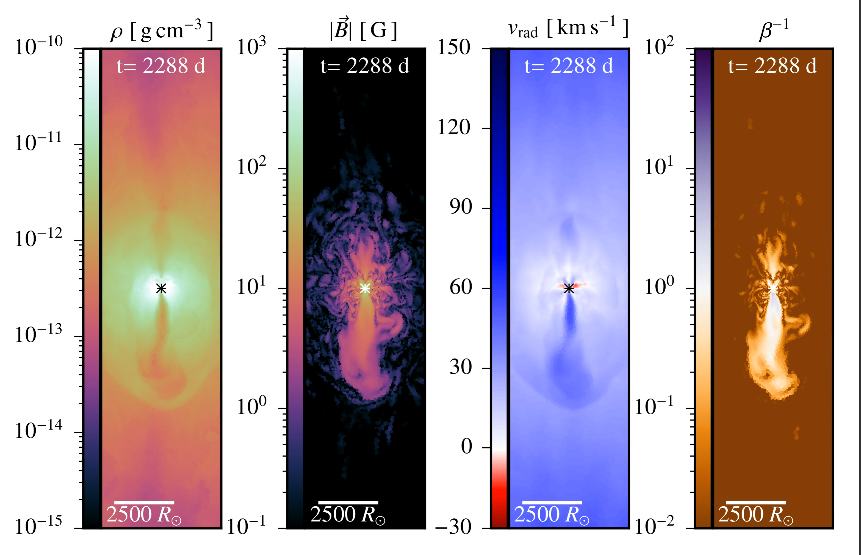}}
	\caption{(\href{https://zenodo.org/record/5596215/files/fig_A3_movie_launching_of_bipolar_outflow.mp4?download=1}{Movie online}) Movie of launching of the bipolar outflow.
Similar to Fig.~\ref{fig:outflow}, we show the evolution of gas density, absolute magnetic field strength $|\vec{B}|$, radial velocity $v_\mathrm{rad}$ and ratio of magnetic to thermal pressure $\beta^{-1}$ perpendicular to the orbital plane.}
	\label{fig:mov-jet}
\end{figure}

Fig.~\ref{fig:mov-rho-face} and Fig.~\ref{fig:mov-rho-edge} illustrate the density evolution in the orbital plane and perpendicular to this plane. Fig.~\ref{fig:mov-jet} shows the launching of the bipolar outflow seen edge-on.

\newpage

\section{Hydrogen and helium recombination} \label{sec:photosphere}

\begin{figure*}
	\centering
    \includegraphics[width=17cm]{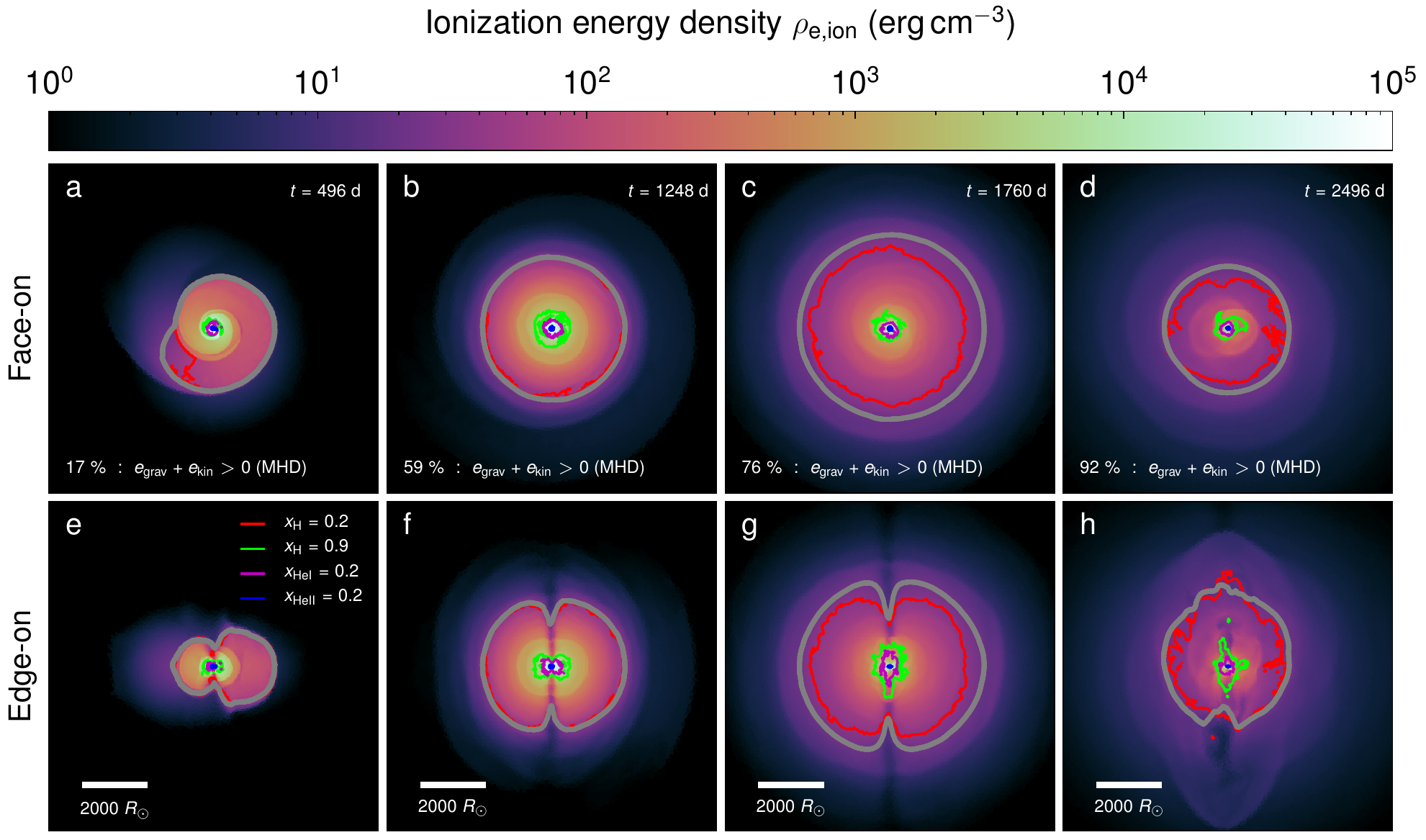}
	\caption{Ionization-energy density in the reference simulation. The ionization-energy density, $\rho_\mathrm{e,ion}$, is shown at four different times in the orbital plane (top panels a--d) and perpendicular to the orbital plane (bottom panels e--h). The grey contours indicate the approximate position of the photosphere at a given time. As marked in the legend of panel (e), the red and green contours enclose a region where the number fraction of ionised hydrogen is between $0.2$ and $0.9$, and thus illustrate the region where hydrogen recombines. The purple and blue contours correspond to HeI and HeII number fractions of $0.2$, \ie helium recombines within these layers, close to the center of the expanding envelope. In panels (a)--(d), the current, unbound mass fractions as measured by the kinetic criterion are given (\cf Fig.~\ref{fig:orbit-and-unbound-mass}).}
	\label{fig:photosphere}
\end{figure*}

In Fig.~\ref{fig:photosphere}, we show the (still) available ionization-energy density at different epochs in the simulation. The photosphere is determined by radially integrating the optical depth $\tau$ from outside in until reaching a value of $\tau=1$. It is illustrated by a grey contour and the grid cells close to the thus-defined photosphere are already relatively large, precluding a precise localisation. The outer edge of the hydrogen recombination front, which we define by a number fraction of ionized hydrogen $x_{\mathrm{H}}$ of less than $0.2$, is located close to the photosphere. However, the inner edge of the hydrogen-recombination region (indicated by $x_{\mathrm{H}} < 0.9$) extends down to the center of the expanding envelope. This shows that energy from recombining hydrogen is mostly released throughout the optically thick region of the envelope and only in minor parts near or outside the photosphere. The outer ionization fronts of HeI and HeII are in even deeper and more opaque layers than those of hydrogen at the center of the envelope.

\end{appendix}

\end{document}